\documentclass[a4paper]{scrartcl} 

\usepackage[a4paper]{geometry}
\usepackage{graphicx}
\usepackage{float}
\usepackage{amssymb}
\usepackage{amsfonts}
\usepackage{amsmath}
\usepackage{cancel}
\usepackage{mathcomp}
\usepackage[ruled,vlined]{algorithm2e}
\usepackage{algorithmic}

\title{Domain Boundary Detection in Hi-C Maps: A Probabilistic Graphical Model Approach}
\author{\parbox{\linewidth}{\centering
Andreas Hofmann, Fatema Zahra Rashid, Fr\'{e}d\'{e}ric Cr\'{e}mazy, Remus T. Dame, Dieter W. Heermann
}}

\date{\today}

\begin{document}

\maketitle

\begin{abstract}
To understand the nature of a cell, one needs to understand the structure of its genome. For this purpose, experimental techniques such as Hi-C detecting chromosomal contacts are used to probe the three-dimensional genomic structure. These experiments yield topological information, consistently showing a hierarchical subdivision of the genome into self-interacting domains across many organisms. Current methods for detecting these domains using the Hi-C contact matrix, i.e. a doubly-stochastic matrix, are mostly based on the assumption that the domains are distinct, thus non-overlapping. For overcoming this simplification and for being able to unravel a possible nested domain structure, we developed a probabilistic graphical model that makes no a priori assumptions on the domain structure. Within this approach, the Hi-C contact matrix is analyzed using an Ising like probabilistic graphical model whose coupling constant is proportional to each lattice point (entry in the contact matrix). The results show clear boundaries between identified domains and the background. These domain boundaries are dependent on the coupling constant, so that one matrix yields several clusters of different sizes, which show the self-interaction of the genome on different scales.

\end{abstract}

\section{Introduction}

Early work using optical microscopy with fluorescent markers established that chromosomes are not randomly organized in the nucleus~\cite{Tark-Dame:2011aa}. Exactly how the chromosomes are organized could not be further revealed by this method, even though multi-color experiments pushed the experimental boundary~\cite{Cass:aa}. At this stage several models have been proposed how the genome is physically organized in space~\cite{Hahnfeldt1993,Yokota1995,sachs95,Yokota1997,bohn10b,heermann11,mirny11}. With the chromosome conformation capture technology (3C)~\cite{dekker03} new data on the organization became available. Whereas the information coming from the microscopy experiments gives a physical relationship between between points in space, i.e., Euclidean distances on single cell data, the 3C (and later the Hi-C data~\cite{dekker09}) yields topological information loosing the embedding into euclidean space, i.e., only neighborhood relationships are revealed attached with a certain probability. Furthermore the information represents an average over many cells. In a way this is very much information one would classify as of mean-field type. Thus the challenge is to develop a model that is consistent with the mean-field result in the sense that it succeeds to re-embed the topological information into euclidean space, i.e., geometrical information and topological information need to be reconciled. 

A crucial part of this process is to identify the structures and substructures that appear in Hi-C heatmap data. Most prominently are the TADs (topologically associated domains). Their defining characteristic is that the interaction frequency within domains is much higher as opposed to that across domains, i.e. the contact matrix resembles a block-diagonal matrix.

There are various different methodological approaches identifying the domain structure in Hi-C contact maps.\\
A first attempt was presented in Dixon et al.~\cite{Dixon_2012} and is based on a two-step strategy. Firstly, the 2D contact information is condensed to the directionality index, a 1D measure encoding both downstream and upstream chromatin interactions. In the second step, a hidden Markov model (HMM) is applied to this data to retrieve the segmentation into domains. Instead of a HMM, it is also possible to translate the directionality index into a test statistics in order to identify significant domain boundaries~\cite{Le_2013}.\\
L\'{e}vy-Leduc et al.~\cite{Levy-Leduc_2014} developed a 2D model that fits a block diagonal matrix to observed contacts using maximum likelihood.\\
Filippova et al.~\cite{Filippova_2014} use dynamic programing to find domains with maximal intra-domain contact frequency.\\
Weinreb et al.~\cite{Weinreb_2016} developed a method to find an optimal TAD hierarchy via dynamic programing.\\
Chen et al.~\cite{Chen_2016} present a method for identifying topological domains based on the spectral decomposition of the graph Laplacian of the Hi-C matrix.


Complementary to the above outlined heuristic and mostly image analysis motivated approaches, one can interpret the Hi-C data as interactions and treat them on this level. Following this idea leads naturally to probabilistic graphical models. In the following section we develop the approach.

\section{Approach}

The main idea is to use an energy based probabilistic graphical model. In fact we will construct a log linear model over a Markov network. For the energy function a possible choice is to use the pair interactions (pairwise node potential) that are defined by the Hi-C heatmap together with feature variables between which the interaction is defined. The energy of the pairs is symmetric. Now rather than learning parameters we sample the feature variables as a function of control parameters. If there is a strong interaction between a pair of nodes than we construct the energy function to favor the feature variables to be similar. On the other hand if there is only a weak interaction between nodes then the feature variables will be uncorrelated. Assume that the feature variables take value $\pm 1$. Within a domain, where the interaction is strong, the feature variables will all have nearly identical average values. Where there is a very weak or no interaction the feature variable will average to zero. Within this scheme domains can be identified by the boundary from values above a certain threshold and zero.

\section{Methods}

Let $\mathbf {C}$ be the matrix containing the raw counts from the Hi-C experiment
\begin{equation}
\mathbf {C} = (c_{ij})_{i,j:1, \dots,n} = 
\begin{pmatrix}
c_{11} & c_{12} & \dots & c_{1n} \\
c_{21} & c_{22} & \dots & c_{2n} \\
\vdots & \vdots & \dots & \vdots \\
c_{n1} & c_{n2} & \dots & c_{nn} 
\end{pmatrix}
\end{equation}

\noindent where $c_{ij}\ge 0$ for  $i,j:1,\dots,n$.
This symmetric non-negative matrix can be normalized~\cite{Sinkhorn:1967aa} such that the row/column sums in the euclidean norm $|| \cdot ||_2$ is one

\begin{equation}
\mathbf {C} \rightarrow \mathbf {\eta} = (\eta_{ij})_{i,j:1,\dots,n} = 
\begin{pmatrix}
\eta_{11} & \eta_{12} & \dots & \eta_{1n} \\
\eta_{21} & \eta_{22} & \dots & \eta_{2n} \\
\vdots & \vdots & \dots & \vdots \\
\eta_{n1} & \eta_{n2} & \dots & \eta_{nn} 
\end{pmatrix}
\end{equation}

i.e., 
\begin{equation}
 1 = \sum_{i=1}^n \eta_{ij} \;\text{for all} \; j:1,\dots,n
\end{equation}

For our approach, the matrix does not need to be doubly stochastic. It rather is convenient to compare later results with respect to the parameters that control coupling strengths. The model and the algorithm presented below just rests on the fact the matrix describes a network where the entries in the matrix represent values for the edges of the network.

\subsection{The Model}

\begin{figure}[H]
\centerline{\includegraphics[width=0.7\textwidth]{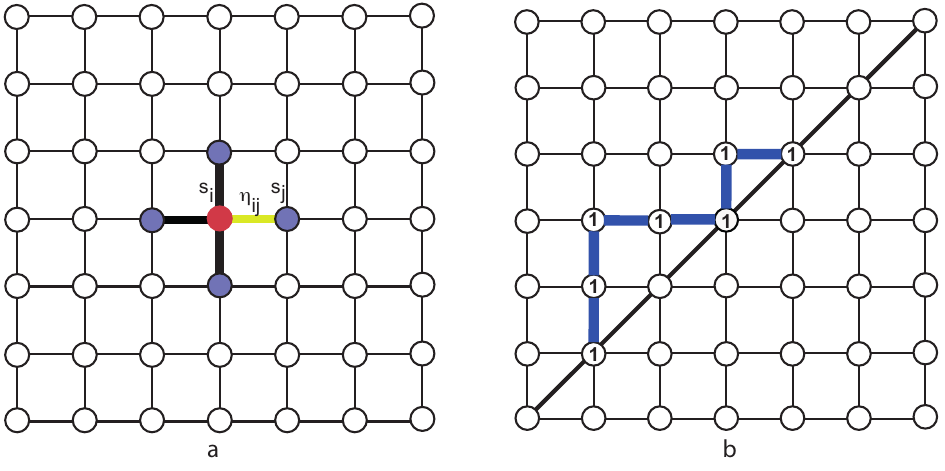}}
\caption{feature lattice carrying the variables $s_i = \pm 1$}\label{fig:01}
\end{figure}

We define feature variables $s_i$ that can take on values $\pm 1$ that are associated with the nodes of a network (of which we have $N=n^2$) which in fact has a simple square lattice structure. The network is defined by the Hi-C matrix presented above where the edges of the network are the entries of the matrix $\eta_{ij}$ and the nodes carry the feature variables. For convenience we restrict ourselves here to just two features. In principle the feature set can be a set $\{0,...q \}$ with  $q\in \mathbf{N}$.   Let $\mathbf {s}=(s_1,...,s_N)$  be a specific feature configuration. Based on the pair-interaction specified by the normalized Hi-C matrix and the feature configuration we specify a symmetric energy function. The idea being that if two nodes (here we restrict ourselves to nearest neighbor nodes) have a high value in the normalized Hi-C matrix then the feature variable should tend to be similar. If the next-nearest neighbor nodes have in turn similar Hi-C entries the feature would be propagated depending on a control parameter that governs the relative strength. The simplest ansatz in this direction is a log-linear model. In this scheme the probability for a specific configuration $\mathbf {s}$ is

\begin{equation}
p(\mathbf {s} | \mathbf {\eta},\alpha,\beta) = \frac{1}{Z}e^{-\epsilon(\mathbf {s},\mathbf {\eta},\alpha,\beta)}
\end{equation}

\noindent with the normalization

\begin{equation}
Z = \sum_{\mathbf {s}} e^{-\epsilon(\mathbf {s},\mathbf {\eta},\alpha,\beta)}
\end{equation}

\noindent and $\epsilon(\mathbf {s},\mathbf {\eta},\alpha,\beta)$ being the energy function. Assuming symmetric pairwise interaction between the nodes with the interaction given by the values of the normalized Hi-C matrix and a possible local bias we use the following form for the energy function

\begin{equation}
\epsilon(\mathbf {s},\mathbf {\eta},\alpha,\beta) = \alpha \sum_{\langle ij \rangle} \eta_{ij}s_i s_j + \beta \sum_i  \eta_{ij}s_i 
\end{equation}

\noindent where $\alpha$ and $\beta$ are control parameters for the strength of the coupling between pairwise nodes ($\alpha$) and $\beta$ controlling the bias. Note that we restrict the pairwise interaction to nearest-neighbor nodes, as depicted in figure~\ref{fig:01}, denoted by the symbol $\langle ij \rangle$.

In the above scheme we are using free boundary conditions. For nodes that have no nearest neighbors in one of the directions that interaction for this direction is taken to be zero.

\subsection{Algorithm}

Given the above model, we sample the feature variables using the Metropolis Monte Carlo method~\cite{Binder-Heermann:1988}. The goal is identify the domains that have on {\em average} equal feature. Since strong interaction favor a like feature, i.e. the Hi-C showed a high probability for the connection between the nodes, with the parameter $\alpha$ we can control the relative strength of the interaction. Since this in turn influences the correlation between the nodes, large values of $\alpha$ will incorporate into domains of like feature also nodes that have a relative lower probability of connectivity. We can thus control how much of a domain structure one wants to explore.

At the start of the algorithm all feature variables are set to $+1$. Because the Hi-C interaction is non-negative, this ensures equilibrium in the sampling using MCMC~\cite{Binder-Heermann:1988}. The sampling is set to last up to a maximum number of iterations or terminates if the moving average of the overall feature variable has changed less than a given value.

To define the border between domains, we use a threshold $c$ above which the average value of the feature at node $i$ belongs to a domain, i.e. we define a characteristic function

\begin{equation}
\chi(i) =     
\begin{cases}
      0, & \text{if}\ \langle s_i\rangle  < c \\
      1, & \text{otherwise}
    \end{cases}
\end{equation}

\noindent where $\langle s_i\rangle$  is the average value of the feature variable $s$. For those nodes that are not strongly connected the average in the MCMC process will tend to zero, whereas those that are strongly connect tend to $+1$ given the initial condition of all nodes having $+1$.

This yields a configuration that has only $0$ or $1$ for each node. To detect the boundaries we delete all nodes where at least one of the nearest neighbor nodes has feature value $0$. Connecting those nodes left hat have characteristic values of $1$ along the diagonal define the border between the domains. With this algorithm, we are able to identify non-rectangular domains as we will shown below. 

\begin{algorithm}[H] 
\caption{Hi-C Domain Structure Identification}
\label{ALGO:Domain}
\begin{algorithmic}
\STATE initialize feature variables with feature $1$
\STATE  $mcs \leftarrow 1$
\WHILE{$mcs < mcsmax$ } 
\STATE generate a realization using Metropolis MC
\STATE compute individual feature average
\STATE  $mcs \leftarrow mcs+1$
\ENDWHILE
\STATE using $\chi$ project the average feature variable to $0$ or $1$
\STATE delete all nodes that have at least one nearest neighbor with $0$
\end{algorithmic}
\end{algorithm}

\subsection{Validation of the Algorithm}

\begin{figure}[ht]
\centerline{\includegraphics[width=0.7\textwidth]{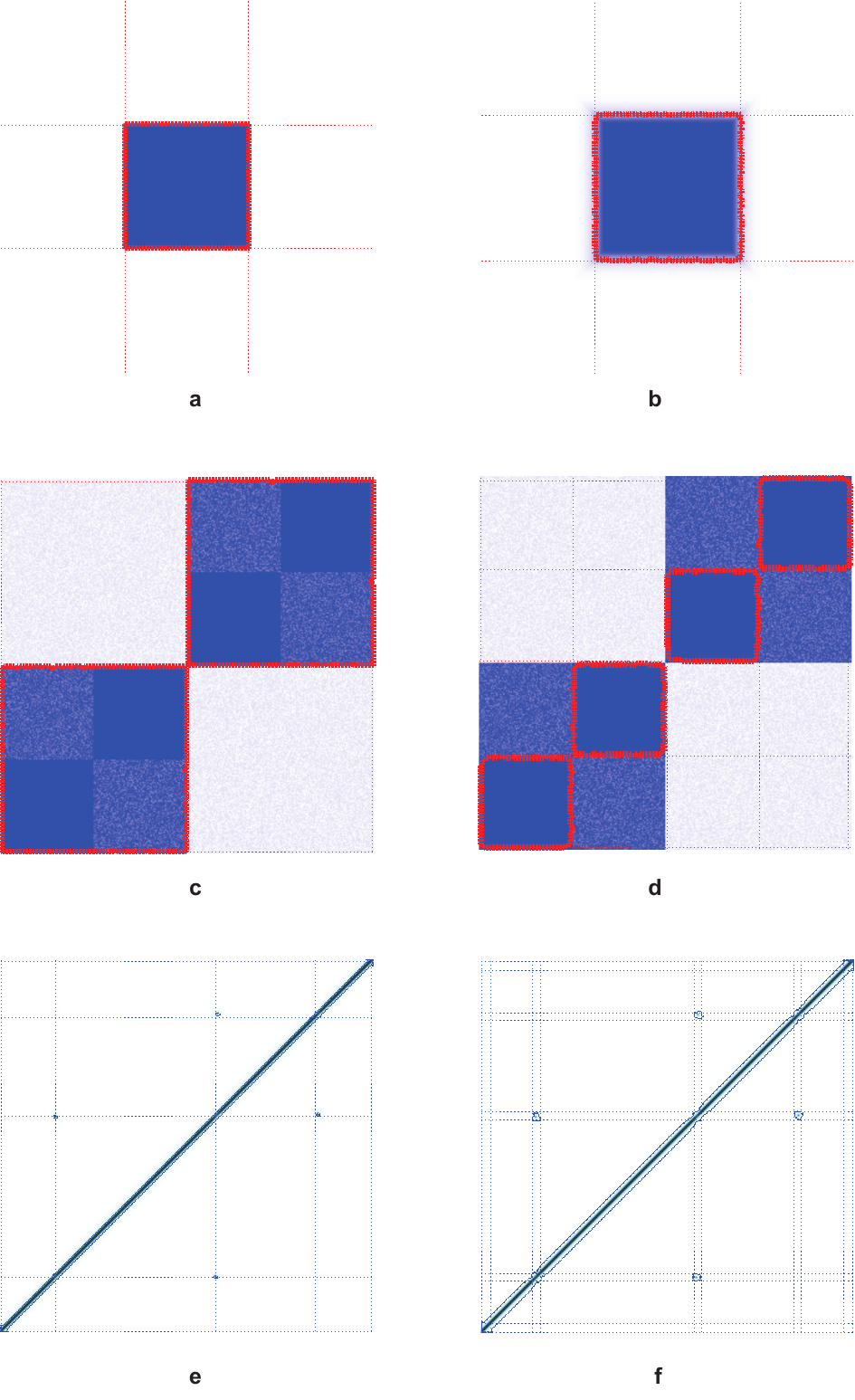}}
\caption{Test cases for the algorithm: a) simple square with no background noise and  $\alpha=10000$, b) simple square with diffuse boundary and no background noise $\alpha=10000$, c) domains within domains with background noise $\alpha=25000$, d) domains within domains with background noise e) and f)  static loop polymer with parameters: $t=1000$ and $t=16000$ respectively. All results are averages over $10000$ Monte Carlo Steps. Dashed lines represent the horizontal (vertical lines identified by the algorithm belonging to a domain.}\label{fig:02}
\end{figure}

The above outlined algorithm was tested using synthetic data. Three cases were considered. First in line is the square domain with sharp and with fuzzy boundary. The result of the domain identification is shown in Figure~\ref{fig:02} (top panel).  In both cases the same control parameters $\alpha = 10000$ and $\beta = 0$ were used. In both cases the square domain is correctly identified. The dashed lines give the horizontal (vertical) identification line of the domain boundary.

The middle panel shows the results of the domain identification against a noisy background and have noise also inside of the domain with varying degree of intensity. In Figure~\ref{fig:02} c the interaction $\beta$ was slightly twice higher than in Figure~\ref{fig:02} d.

The bottom panel shows that also non-square domains can be identified which are associated with loops in the chromosome conformations.

\section{Results: Domain Detection within Hi-C contact maps}

As discussed in the introduction, existing methods for domain detection assume that the domains are distinct, contiguous blocks of increased contact probability. In order to detect both domains and more complex structures including loops or loop domains without bias, we developed our probabilistic graphical model that makes no a priori assumptions on the domain structure. Since our approach is relying on the graph theoretic interpretation of Hi-C matrices, it does not yield domain boundaries, but rather a contour separating contact probabilities of an adjustable strength from the background. This iso-strength contour allows identification and characterization of compartments irrespective of whether or not there are contiguous domains in the form of squares.

\begin{figure}[H]
\centering
\hskip-0.97\textwidth\large\textsf{A}\\
\includegraphics[width=0.99\textwidth]{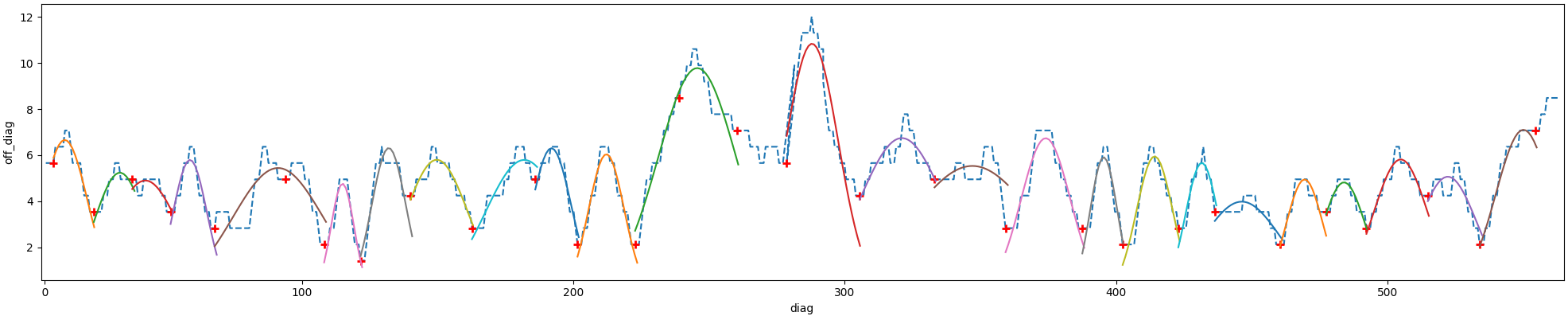}\\
\hskip-0.97\textwidth\large\textsf{B}\\
\includegraphics[width=0.99\textwidth]{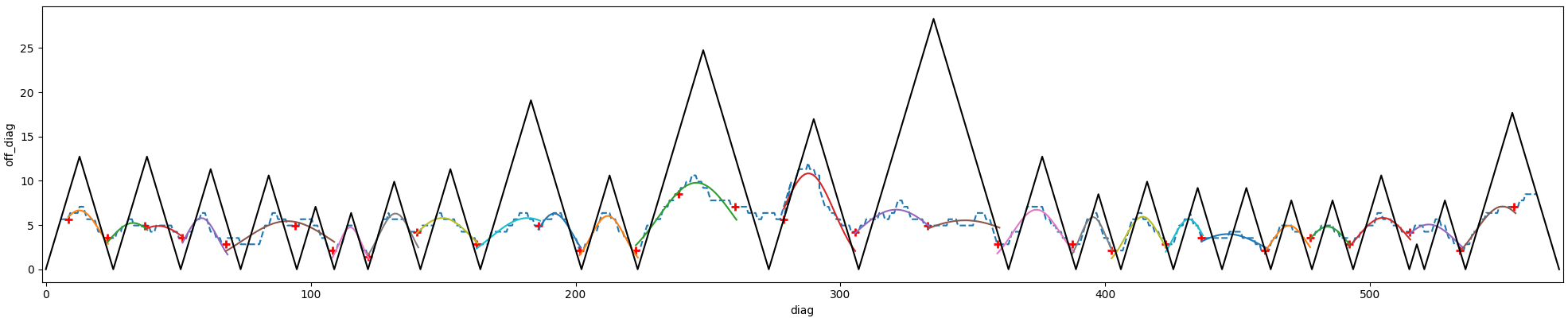}\\
\hskip-0.97\textwidth\large\textsf{C}\\
\hspace{-0.12cm}\includegraphics[width=0.997\textwidth]{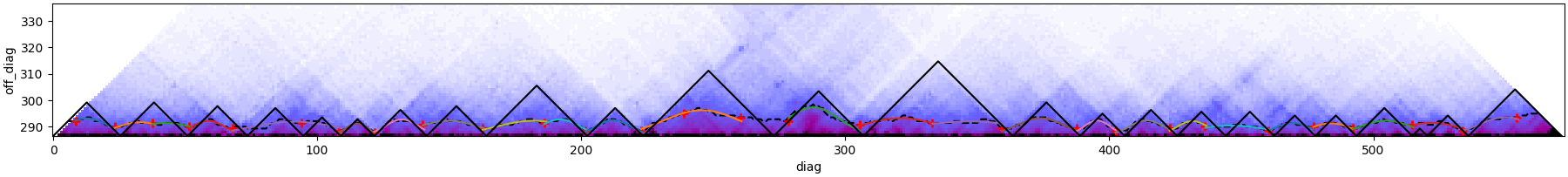}
\caption{Excerpts of the 45° anti-clockwise rotated Hi-C contact maps of the C. crescentus chromosome and results of both the approach using the directionality index and our probabilistic graphical model. Contrary to the directionality index approach our method does not yield domain boundaries or rather a domain structure (black), but a contour (dashed blue) separating contact probabilities of a certain strength (depending on the coupling constant) from the background. This iso-strength contour allows to characterize the compartmentalization irrespective of whether or not there are contiguous domains in the form of squares.
}\label{fig:03}
\end{figure}

In order to illustrate and visually compare the results of our approach and that based on the directionality index, we used the Hi-C contact maps of C. crescentus. The contiguous domain structure computed on the basis of the directionality index as well as the iso-strength contour yielded by our method permit characterization of the compartmentalization of the wild-type C. crescentus chromosome (see Figure~\ref{fig:03}). 

\section{Discussion}

We have developed a probabilistic graphical model to study the domains structure visible in Hi-C heat maps. This model is based on a symmetric energy model where the interaction parameters come from the normalized entries of the heat map. Here the heat map is interpreted as a graph with $N=n^2$ nodes each node having a feature variable. Already a model where the feature variable has just two values is sufficient to identify synthetic domains. This domains incorporate partial noise as would be expected to the noise in the actual heat maps. The domains themselves are set against a background of noise. The model is able to identify the noise through the average feature variable which is clearly distinct to the one in the domain. Within the domain, depending on the strength of the control parameter, the average value of the feature variable is homogeneous. This leads to the clear identification of the domain boundary as those nodes that have at least one of the nearest neighbors having a feature value different from the others.

\vspace{0.5cm}

\noindent \textbf{Acknowledgments} \newline
We would like to thank Remus Dame for the very stimulating discussions.

\vspace{0.5cm}

\noindent \textbf{Funding} \newline HFSP Leiden / Birmingham / Heidelberg as well as the Heidelberg Graduate School of Mathematical and Computational Methods in the Sciences (HGSMathComp)


\bibliographystyle{unsrt}
\bibliography{hic-aps}

\end{document}